\documentclass[aps,prd,amssymb,amsmath,amsfonts,superscriptaddress,nofootinbib,eqsecnum,reprint,showpacs,longbibliography]{revtex4-2}

\usepackage{graphicx}
\usepackage{lmodern}
\usepackage{amsmath,amssymb}
\usepackage{mathrsfs}
\usepackage{amsfonts}
\usepackage[utf8]{inputenc}
\usepackage{url}
\usepackage[colorlinks]{hyperref}
\usepackage{xcolor}
\usepackage[normalem]{ulem}
\usepackage{orcidlink}
\usepackage{placeins}
\usepackage{mathtools}
\usepackage{rotating}
\usepackage{tabularx}
\usepackage{enumitem} 
\usepackage{xspace}
\usepackage{subcaption}
\usepackage{multirow}
\usepackage{booktabs}
\usepackage[justification=justified,singlelinecheck=false]{caption}
\usepackage{longtable}
\usepackage{ltcaption}
\makeatletter
\expandafter\let\csname longtable*\endcsname\relax
\expandafter\let\csname endlongtable*\endcsname\relax
\makeatother

\renewcommand{\arraystretch}{1.75}

\definecolor{dodgerblue}{HTML}{1E90FF}
\definecolor{NatalieGreen}{HTML}{336600}
\definecolor{TimPurple}{HTML}{8303a6}

\hypersetup{
     colorlinks=true,
     linkcolor=dodgerblue,
     filecolor=dodgerblue,
     citecolor = dodgerblue,      
     urlcolor=dodgerblue,
     }

\newcommand{\Potsdam}{Universit\"at Potsdam, Institut f\"ur Physik und Astronomie, \\
Haus 28, Karl-Liebknecht-Str. 24/25, 14476, Potsdam, Germany}

\begin{abstract}
Gravitational-wave observations of binary neutron star mergers probe the neutron star equation of state through finite-size effects on the waveform. These include the tidal deformation of the stars due to their respective companion, but  for spinning neutron stars also include the spin-induced quadrupole moment of the stars. Typically, to avoid introducing additional equation of state dependent parameters, waveform models relate this quadrupole moment to the tidal deformability through so-called quasi-universal relations. These are derived under the slow-rotation approximation, which at larger spins, i.e., for millisecond pulsars, becomes unreliable. We attempt to extend the existing quasi-universal relations by explicitly accounting for the spin, calibrating it against numerically computed quadrupole moments across a broad set of equations of state and spins $0.15 \leq |\chi| \leq 0.6$ with the $\chi\rightarrow 0$ limit enforced by construction. The resulting spin-extended relation reduces the median absolute percentage error relative to the slow-rotation relation by 66\% across this range. Parameter estimation using the spin-extended relation reduces biases in recovered intrinsic parameters in some cases, though biases from mass-spin degeneracies dominate largely when a population is considered. These results provide a more accurate description of the spin-induced quadrupole moment for spinning neutron star binaries, relevant for waveform modelling with third-generation gravitational-wave detectors.

\end{abstract}

\begin{document}

\title{Extending the Love-Q Relation to Rapidly Rotating Neutron Stars for Gravitational-Waveform Modelling}

\author{Natalie Williams \orcidlink{0000-0002-5656-8119}} 
\affiliation{\Potsdam}

\author{Tim Dietrich \orcidlink{0000-0003-2374-307X}} 
\affiliation{\Potsdam}
\affiliation{Max Planck Institute for Gravitational Physics (Albert Einstein Institute), \\ Am M\"uhlenberg 1, Potsdam 14476, Germany}

\date{\today}

\maketitle

\section{Introduction}
Binary neutron star (BNS) mergers observed via gravitational waves (GWs) provide a unique probe of the neutron star (NS) equation of state (EOS), through the imprint of finite-size effects on the GW waveform. The dominant matter effect is captured with the quadrupolar tidal deformability $\Lambda$, which characterizes the induced quadrupole response of a NS to its companion's tidal field. To date, GW observations of BNS systems remain limited, with only GW170817~\cite{LIGOScientific:2017vwq, LIGOScientific:2017zic, LIGOScientific:2017ync} detected with Advanced LIGO~\cite{LIGOScientific:2014pky} and Advanced Virgo~\cite{VIRGO:2014yos} yielding significant constraints on tidal effects. However, the advent of third-generation (3G) detectors such as Einstein Telescope (ET)~\cite{Punturo:2010zz, Branchesi:2023mws, ET:2019dnz, ET:2025xjr} and Cosmic Explorer~\cite{Evans:2021gyd, Reitze:2019iox} will substantially increase both the number and signal-to-noise ratio (SNR) of BNS observations, enabling substantially tighter EOS constraints while placing correspondingly stronger requirements on waveform accuracy.

In addition to the tidal deformability, rotation of NSs introduces an additional finite-size effect through the spin-induced quadrupole moment (SIQM), arising from the rotational deformation of the NS. This effect is captured within waveform models through the spin-induced quadrupolar deformability $C_Q$ which is related to the SIQM $Q = -C_Qm^3\chi^2$ ~\cite{, Laarakkers:1997hb}, where $m$ is the NS mass. Unlike the tidal deformability, which first enters the waveform effectively at 5PN order, the SIQM contributes at 2PN order, making it particularly relevant for 3G detectors~\cite{Harry:2018hke} with low-frequency sensitivity extending to $\sim$5\,Hz.

Electromagnetic observations of galactic BNS systems suggest that NSs in merging binaries are typically slowly rotating; the double pulsar PSR J0737-3039A is expected to have a dimensionless spin of $|\chi| \sim 0.04$ at merger~\cite{Burgay:2003jj}. However, there is observational evidence for significant spins in NS systems: PSR J1807-2500B has an inferred upper limit of $|\chi| \lesssim 0.2$~\cite{Lynch_2012}, while the fastest spinning pulsar has $|\chi| \lesssim 0.4$~\cite{Hessels:2006ze}. Furthermore, 3G detectors will observe a much larger and more distant BNS population than is currently accessible electromagnetically. If present, such spins would offer a valuable insight into BNS formation channels and the properties of the post-merger remnant.

Similarly to the tidal deformability, the SIQM is an EOS-dependent quantity. Rather than increase the dimensionality of waveform models by treating $C_Q$ as an independent parameter, current waveform models instead use \textit{quasi-universal relations} (qURs), which exploit approximate EOS-independence to map between NS quantities. In relation to the SIQM, a qUR linking $\Lambda$ and $C_Q$ is used, allowing waveform models to include the SIQM information without introducing additional independent EOS-dependent parameters~\cite{Yagi:2016bkt}.

Crucially, however, the qUR used to obtain $C_Q$ is derived under the slow-rotation approximation, and its accuracy is therefore expected to degrade with increasing spin~\cite{Doneva:2013rha, Yagi:2014bxa}. For current detector sensitivities, this assumption is likely sufficient. However, as 3G detectors increase overall measurement precision, even subdominant effects become increasingly important, making a quantitative assessment of the accuracy of qURs essential. Ref.~\cite{Williams:2026jqv} showed that applying the current $C_Q$ qUR to extreme EOSs introduces biases in inferred parameters at spins as low as $|\chi| = 0.15$. Despite this being very high in relation to the observed spins of electromagnetically observed NSs, is well within the astrophysically relevant $|\chi| \lesssim 0.4$ range, that current GW  parameter estimation for BNSs also samples over with a high-spin prior, and therefore where the model is assumed to remain valid.

In this work, we extend the current qUR to account for dependence on spin, fitting to numerically calculated $C_Q$ across a wide set of EOSs for spins $|\chi| \leq 0.6$.  We quantify the accuracy of this extended relation relative to the existing slow-rotation qUR across this EOS set and a range of spin magnitudes, and assess its impact on parameter inference for spinning BNS systems in the 3G era.

This paper is organised as follows: In Sec.~\ref{Sec:construction} we describe the methodology, including the EOS dataset used and the fitting of the spin-extended qUR. In Sec.~\ref{Sec:validation} we validate the qUR. In Sec.~\ref{sec:waveform} we explore the impact of the spin-extended qUR on the waveform and in Sec.~\ref{Sec:PE} on parameter estimation. Finally in Sec.~\ref{Sec:conclusion} we conclude. Throughout the work, we set $G = c = 1$. For the individual component masses we define $m_A$ and $m_B$ as the primary and secondary masses respectively, where these subscripts carry forward onto other NS parameters. The mass ratio is taken as $q = m_B/m_A \leq 1$, and aligned spin components are given by $\chi_A$ and $\chi_B$.

\section{Methodology}
\label{Sec:construction}
\subsection{Equations of State}
\begin{figure}
    \centering
    \captionsetup{justification=raggedright,singlelinecheck=false}
    \includegraphics[width=0.9\linewidth]{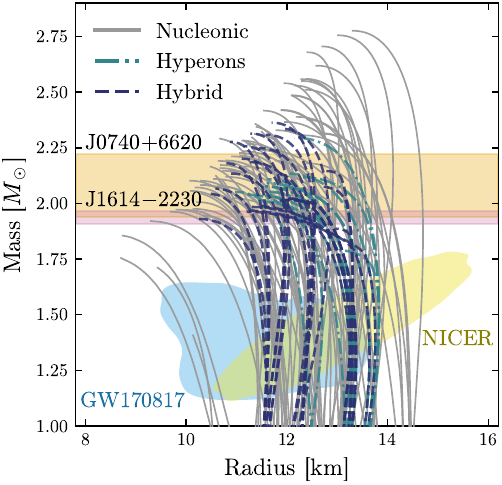}
\caption{Mass-radius curves for the 128 tabulated EOSs used in the calibration of the spin-extended qUR, coloured by particle content: nucleonic (grey), hyperons (teal) and hybrid (quark-hadron) (blue). Shown alongside are the $2\sigma$ measured masses of the millisecond pulsars PSR~J0740$+$6620~\cite{Fonseca:2021wxt} (orange band) and PSR~J1614$-$2230~\cite{NANOGrav:2023hde, Riley:2019yda} (red band), the $90\%$ credible region from the low-spin \textsc{PhenomPNRT} analysis of GW170817~\cite{LIGOScientific:2018hze, LIGOScientific:2017vwq, LIGOScientific:2018hze, LIGOScientific:2019lzm, LIGOScientific:2018cki, LIGOScientific:P1800115} (blue region), and the $90\%$ credible region reported by NICER for PSR~J0030$+$0451~\cite{Miller:2019cac} (yellow region).}
    \label{fig:EOS}
\end{figure}

In the fitting set of the spin-extended qUR, we include 128 EOSs from the CompOSE~\cite{CompOSE, Typel:2013rza, Oertel:2016bki, CompOSECoreTeam:2022ddl} database, an online open-access database of tabulated EOSs. This dataset is detailed in App.~\ref{Appdx:EOS}, and shown in Fig.~\ref{fig:EOS}. We include all currently available cold NS EOSs with the exception of one which contained too few datapoints for our pipeline.  This dataset deliberately includes hadronic, hyperonic, and
hybrid models. We also do not make any requirements to align with observational constraints in order to remain as agnostic as possible.

\begin{figure}
    \centering
    \captionsetup{justification=raggedright,singlelinecheck=false}
    \includegraphics[width=1\linewidth]{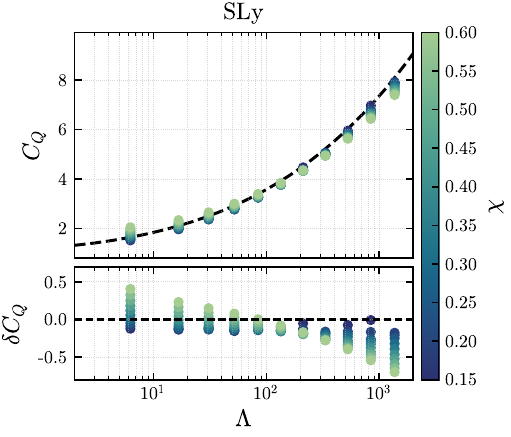}
\caption{Spin-induced quadrupolar deformability $C_Q$ obtained with \texttt{rns} for the SLy EOS, shown as a function of the tidal deformability $\Lambda$ and coloured by the dimensionless spin $\chi$. \textit{Upper panel:} $C_Q$ from \texttt{rns} (points) alongside the slow-rotation qUR of Eq.~\eqref{eq:slowrot-qUR} (black dashed line). \textit{Lower panel:} residual $\delta C_Q$ of the \texttt{rns} data with respect to the slow-rotation qUR, with the zero line marked (black dashed).}
    \label{fig:SLY-SIQM}
\end{figure}

\begin{figure*}[t!]
    \centering
\captionsetup{justification=raggedright,singlelinecheck=false}
    \includegraphics[width=0.8\linewidth]{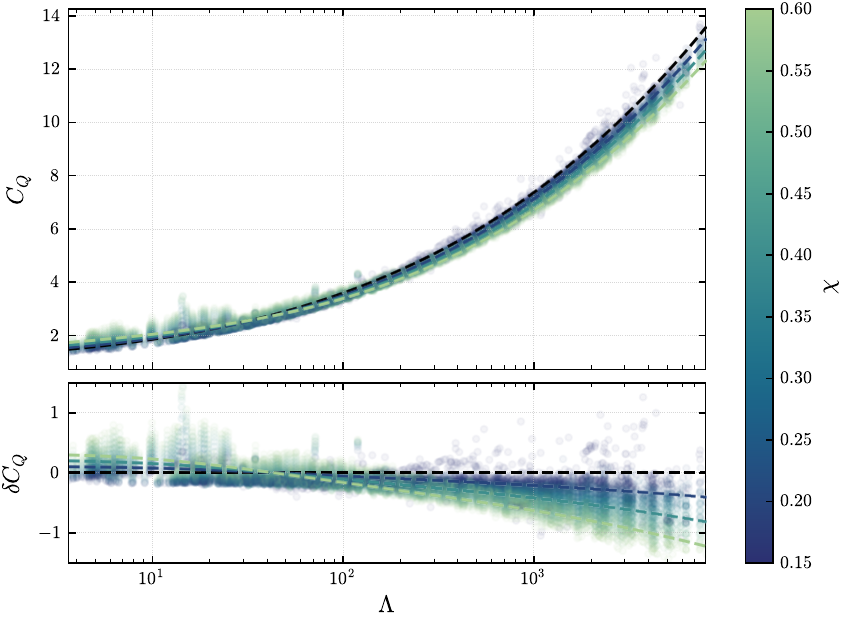}
\caption{As Fig.~\ref{fig:SLY-SIQM}, but for all 128 EOSs of Table~\ref{tab:eos_summary}. \textit{Upper panel:} $C_Q$ from \texttt{rns} (points) alongside the slow-rotation qUR (black dashed line) and the spin-extended qUR of Eq.~\eqref{eq:spin-qUR} for fixed $\chi$ (coloured dashed lines). \textit{Lower panel:} residual $\delta C_Q$ of the \texttt{rns} data with respect to the slow-rotation qUR (points), with the corresponding prediction of the spin-extended qUR overlaid (coloured dashed lines) and the zero line marked (black dashed).}
    \label{fig:fit}
\end{figure*}

\subsection{Computing $C_Q$}
We numerically solve the Einstein equations coupled to the equations of hydrostatic equilibrium for uniformly rotating, perfect-fluid NSs with the Rapidly Rotating Neutron Star \texttt{rns} code~\cite{Stergioulas:1994ea, Stergioulas:2003yp}, which assumes stationarity, axial symmetry about the rotation axis, and
reflection symmetry about the equatorial plane. For each EOS we calculate $C_Q$ across 12 uniformly spaced (gravitational) masses $m\in[1M_{\odot}, M_{\rm TOV}]$ and 10 uniformly spaced spins $\chi \in [0.15, 0.6]$, where $M_{\rm TOV}$ is the maximum TOV mass, giving 120 $C_Q$ values for each EOS. When using \texttt{rns} we use a grid of $151 \times 301$ points in $(\mu, s)$, where $\mu = \cos\theta$ and $s = r/(r + r_{\mathrm{e}})$ is a compactified radial coordinate with $r_{\mathrm{e}}$ the coordinate equatorial radius. This high resolution is required to accurately recover $C_Q$. For a given spin and gravitational mass, we use the \textsc{-t jmoment} method, which calculates an array of NS properties, given an angular momentum and central density. We then change the central density iteratively until the output converges on the target mass. We find that at smaller spins $\chi < 0.15$ the solver does not converge on a reliable solution. We demonstrate the deviation of the rapidly rotating solution from the slow rotation approximation in Fig.~\ref{fig:SLY-SIQM} for the EOS SLy~\cite{Danielewicz:2008cm, Gulminelli:2015csa, Chabanat:1997un}. The residuals are defined as $\delta C_Q = C_Q^{\mathrm{rns}} - C_Q^{\mathrm{SR}}$, where $ C_Q^{\mathrm{rns}}$ is obtained from \texttt{rns} and $C_Q^{\mathrm{SR}}$ is obtained from the slow-rotation qUR. Across $\Lambda \gtrsim 100$, the slow-rotation qUR systematically overestimates $C_Q$ for rapidly rotating NSs. This can be understood as a consequence of the approximation retaining the non-rotating moment of inertia, whereas centrifugal support in a rapidly rotating star yields a higher moment of inertia. At fixed $\chi$, and hence fixed angular momentum, this corresponds to an overestimate of the angular velocity and thus of the rotational deformation, such that $C_Q$ is overpredicted. At low tidal deformabilities $\Lambda \lesssim 100$, the residuals change sign past a turning point, with the slow-rotation qUR instead underpredicting $C_Q$ for all EOSs. This arises because each rotating configuration is compared against the relation evaluated at the tidal deformability of the non-rotating star of the same gravitational mass. Since
rotation increases the mass, the rotating star corresponds structurally to a lighter, less compact configuration of larger $\Lambda$, and evaluating the increasing function $C_Q(\Lambda)$ at the smaller value underpredicts $C_Q$. The size of this offset grows towards $M_{\mathrm{TOV}}$, and hence dominates at low $\Lambda$. We confirm this by checking that when comparing $C_Q$ to compactness, the turning point is retained when the non-rotating value is used but absent when the rotating value is used. We nonetheless retain $\Lambda$, as this is the parametrisation used in current waveform models, and incorporate this phenomenology into the fit.

\begin{figure*}[t]
    \centering
    \captionsetup{justification=raggedright,singlelinecheck=false}
    \includegraphics[width=1\linewidth]{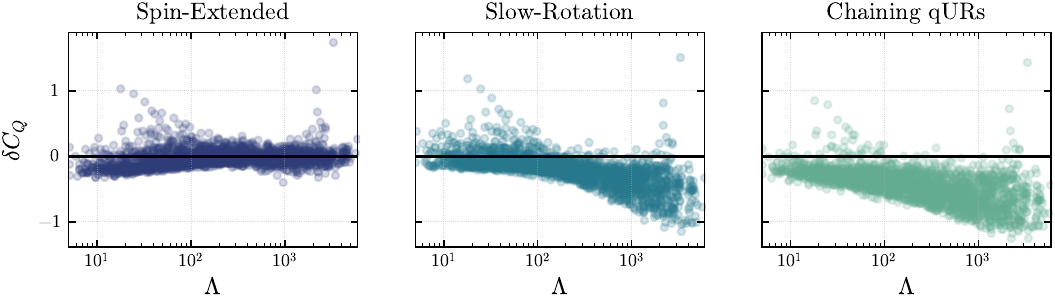}
    \caption{Residuals $\delta C_Q$ between the validation dataset and three models for $C_Q$, as a function of $\Lambda$. \textit{Left panel:} the spin-extended qUR of this work (blue), with median absolute percentage error $1.66\%$. \textit{Centre panel:} the slow-rotation qUR (teal), with median absolute percentage error $4.88\%$. \textit{Right panel:} existing qURs from the literature chained together and inserted into the rapid-rotation $I$--$Q$ relation of Chakrabarti et al.~\cite{Chakrabarti:2013tca} (green), with median absolute percentage error $10.02\%$. The zero line is marked in each panel (black solid).}
    \label{fig:validation}
\end{figure*}

\subsection{Fitting Procedure}
We show the calculated $C_Q$ from all EOSs in Fig.~\ref{fig:fit}, alongside the deviation from the slow rotation approximation qUR. We find empirically that the best fit is given by
\begin{equation}
    C_Q(\Lambda, \chi) = C_Q^{\rm SR}(\Lambda) \times \bigg[ 1 + \frac{|\chi|(a_1 + a_2\Lambda)}{1 +a_3 \Lambda} \bigg]
    \label{eq:spin-qUR}
\end{equation}
with fitting parameters $a_1 = 0.455414$, $a_2 = -0.0107717$, $a_3 = 0.0714282$ obtained by nonlinear least-squares regression against the full set of $15\,360$ $C_Q$ values. A wide selection of functional forms were considered including higher-order polynomials in $\chi$ and $\ln\Lambda$, Pad\'e approximants, forms even in $\chi$ as expected from a slow-rotation expansion~\cite{Yagi:2014bxa}, and fits performed in logarithmic and exponential space, though we found the form of Eq.~\eqref{eq:spin-qUR}, linear in $\chi$, to give best agreement with the data. This ansatz also accounts for the turning point, which we find to be at $\Lambda\sim42$. By construction, in the limit of $\chi \rightarrow 0$, it reduces to the slow-rotation approximation qUR as required~\cite{Yagi:2016bkt}
\begin{equation}
\ln C_Q^{\rm SR}(\Lambda) = \sum_{i=0}^{4}b_i (\ln \Lambda)^i
    \label{eq:slowrot-qUR}
\end{equation}
where $b_0 = 0.1940, b_1 = 0.09163, b_2 = 0.04812, b_3 = - 0.004283, b_4 = 0.00012450$. The slow-rotation qUR alongside the spin-extended qUR is shown in Fig.~\ref{fig:fit}. The rational nature of the spin dependent qUR function results in sensible extrapolation at higher tidal deformabilities. We also note that the region $|\chi| < 0.15$ lies outside our calibration range and Eq.~\eqref{eq:spin-qUR}
should be regarded as an extrapolation there bounded by the slow-rotation limit.

\section{Validation}
\label{Sec:validation}
We validate the fit against data not used within the fitting, generating 100 values of $C_Q$ for each EOS from the same ranges as previous, leading to a total of 12\,800 datapoints. In Fig.~\ref{fig:validation} we show the residual between the validation dataset and the spin-extended qUR which has a median absolute percentage error of 1.66\%, next to the residual from the slow-rotation qUR with a median absolute percentage error of 4.88\%, corresponding to a 66\% reduction in error. 

As an independent literature-based comparison, we also constructed
$C_Q(\Lambda,\chi)$ by chaining together existing qURs~\cite{Haberland:2025luz}. Starting from $\Lambda$, the
non-rotating compactness and moment of inertia are obtained from the
$C$-Love and $I$-Love relations of Yagi and Yunes~\cite{Yagi:2016bkt}. These in turn set the Keplerian angular
velocity $\Omega_K$ and Keplerian-limit moment of inertia $I_K$ via the
relations of Konstantinou and Morsink~\cite{Konstantinou:2022vkr} and
Kr\"uger and V\"olkel~\cite{Kruger:2023olj}. Interpolating $I(\Omega)$
between its non-rotating and Keplerian values and requiring consistency
with $\chi$ fixes the spin frequency and hence the dimensionless moment of inertia $\bar{I}$, which
is finally mapped to $C_Q$ by inverting the rapid-rotation $I$--$Q$
relation of Chakrabarti et al.~\cite{Chakrabarti:2013tca}, evaluated at
fixed $\chi$. The resulting estimate (Fig.~\ref{fig:validation}, right
panel) deviates from the validation data considerably, with a median absolute percentage error of 10.02\%. This is expected: each relation in the chain is
calibrated against directly-computed stellar quantities, not against the
qUR-derived output of a preceding step, so composing them propagates and
compounds their individual errors well beyond the regime
in which any single relation was validated. We show these however to demonstrate that the spin-extended qUR from this work stands independently of other rapidly rotating qURs from the literature.

\section{Impact on the Waveform}
\label{sec:waveform}

\subsection{GW dephasing}
 We next explore how the spin-extended qUR impacts the GW waveform. The SIQM  enters the BNS self-spin phase at 2PN order for NS $A$ and $B$
\begin{align}
\psi_{\mathrm{SS}} &= \frac{3x^{-5/2}}{128\nu}\left( \psi^{A}_{\mathrm{SS,2PN}}x^2 + \psi^{A}_{\mathrm{SS,3PN}}x^3 + \psi^{A}_{\mathrm{SS,3.5PN}} x^{7/2} \right) \notag \\
&\quad + [A\leftrightarrow B],
\label{eq:spinphase}
\end{align}
with~\cite{Poisson:1997ha}
\begin{align}
\psi^{A}_{\text{SS,2PN}} &= -50\hat{C}^A_Q X_A^2\chi_A^2, \notag \\[4pt]
\psi^{A}_{\text{SS,3PN}} &= \frac{5}{84}\left(9407 + 8218X_A - 2016X_A^2\right) \notag \\
&\quad \times \hat{C}^A_Q X_A^2\chi_A^2, \notag \\[4pt]
\psi^{A}_{\text{SS,3.5PN}} &= -400\pi\hat{C}^A_Q X_A^2\chi_A^2,
\label{eq:sscoeffs}
\end{align}
where $\hat{C}^A_Q = C^A_Q -1$ to account for the point particle contribution already present in the binary black hole phase. The spin cubed term is given by
\begin{align}
\psi^{A}_{S^3,3.5\text{PN}} = \frac{3x^{-5/2}}{128\nu}\Bigg\{10\Bigg[\left(X_A^2 + \frac{308}{3}X_A\right)\chi_A \notag \\
+ \left(X_B^2 - \frac{89}{3}X_B\right)\chi_B\Bigg]\hat{C}^A_Q X_A^2\chi_A^2 \notag \\
- 440\hat{C}^A_{\text{Oc}}X_A^3\chi_A^3\Bigg\}x^{7/2} + [A \leftrightarrow B],
\end{align}
where likewise $\hat{C}^A_{\text{Oc}} = C^A_{\text{Oc}} -1$.

The corresponding qUR for $C_{\rm Oc}$ is given in \cite{Abac:2023ujg}, which also assumes the slow-rotation approximation, and we use it throughout this work, so as to focus only on the leading order quadrupole effect of $C_{\rm Q}$.

\begin{figure}
    \centering
    \captionsetup{justification=raggedright,singlelinecheck=false}
    \includegraphics[width=1\linewidth]{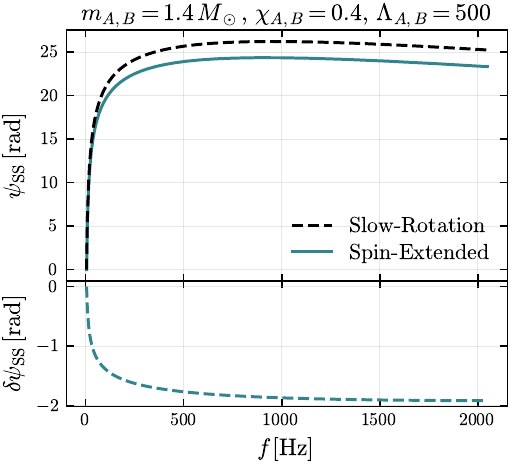}
    \caption{Self-spin contribution to the GW phase, $\psi_{\rm SS}$, for an equal-mass system with $m_{A,B}=1.4\,M_\odot$, $\chi_{A,B}=0.4$ and $\Lambda_{A,B}=500$, aligned at $5\,$Hz. \textit{Upper panel:} $\psi_{\rm SS}$ when using the spin-extended qUR (teal solid) and the slow-rotation qUR (black dashed). \textit{Lower panel:} the phase difference $\delta\psi_{\rm SS}$ between the two (teal dashed).}
    \label{fig:dephasing}
\end{figure}

Figure~\ref{fig:dephasing} shows the self-spin phase generated from 5\,Hz when using the slow-rotation qUR, compared to using the spin-extended qUR for a generic binary system. As a 2PN effect at leading-order, the dephasing is most prominent at low frequencies. For this specific equal-mass system $m = 1.4 M_{\odot}$ with moderate tides $\Lambda=500$ and high spin $\chi = 0.4$, a phase shift of up to $\sim$ 2 rad is obtained. This represents a small, but non-negligible difference within the waveform given the sensitivity of 3G GW detectors. 

\begin{figure}
    \centering
    \captionsetup{justification=raggedright,singlelinecheck=false}
    \includegraphics[width=1\linewidth]{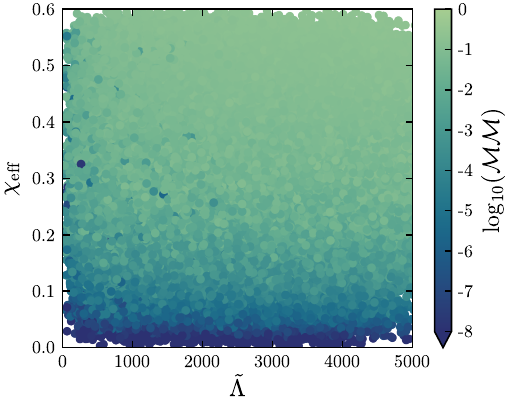}
    \caption{Mismatches $\mathcal{MM}$ between waveforms generated with the slow-rotation qUR and with the spin-extended qUR, shown in the plane of effective spin $\chi_{\rm eff}$ and effective tidal deformability $\tilde{\Lambda}$ and coloured by $\log_{10}\mathcal{MM}$.}
    \label{fig:mismatches}
\end{figure}

\begin{figure*}[!ht]
\captionsetup{justification=raggedright,singlelinecheck=false}
    \centering

        \centering
        \includegraphics[width=0.8\linewidth]{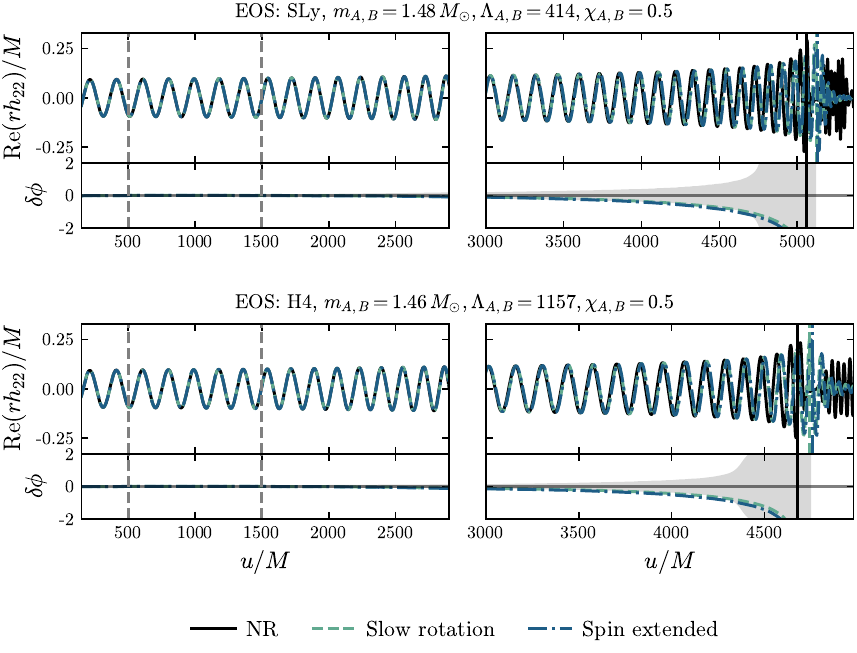}
        \label{fig:NR_SLy}
    \hfill
\caption{Comparison of \textsc{IMRPhenomXAS\_NRTidalv3} waveforms against \texttt{BAM} numerical-relativity data, for two equal-mass, equal-spin ($\chi_{A,B} = 0.5$) binaries: the SLy EOS with $m_{A,B} = 1.48\,M_\odot$ and $\Lambda_{A,B} = 414$ (upper), and the H4 EOS with $m_{A,B} = 1.46\,M_\odot$ and $\Lambda_{A,B} = 1157$ (lower), as labelled. \textit{Upper panels:} real part of the $(2,2)$ strain mode $\mathrm{Re}(rh_{22})/M$ as a function of retarded time $u/M$, for the numerical-relativity data (black solid) and for the waveform model using the slow-rotation qUR (green dashed) and the spin-extended qUR (blue dash-dotted). \textit{Lower panels:} time-domain phase difference $\delta\phi$ between the numerical-relativity data and each waveform model, with the numerical-relativity error shown as the grey shaded band, and computed from the two highest resolutions. The left and right columns show the early inspiral and the approach to merger respectively; the alignment window $u/M \in [500,1500]$ (grey dashed vertical lines) and the merger time (black solid vertical line) are marked.}
    \label{fig:NR}
\end{figure*}
\subsection{Mismatches}

To quantify the disagreement across the parameter space between waveforms using the slow-rotation qUR $h^{\mathrm{SR}}$ and the spin-extended qUR $h^{\mathrm{SE}}$, we compute the mismatch: 
\begin{equation}
\label{eq:MM}
    \mathcal{MM}(h^{\mathrm{SR}}, h^{\mathrm{SE}}) = 1 - \mathrm{max}_{\phi_c, t_c} \frac{\langle h^{\mathrm{SR}}(\phi_c, t_c)|h^{\mathrm{SE}}\rangle}{\sqrt{\langle h^{\mathrm{SR}}|h^{\mathrm{SR}}\rangle\langle h^{\mathrm{SE}}|h^{\mathrm{SE}}\rangle}},
\end{equation}
with the noise weighted inner product being
\begin{equation}
    \langle h^{\mathrm{SR}}|h^{\mathrm{SE}}\rangle = 4 
  \ \mathrm{Re}\int^{f_{\mathrm{max}}}_{f_{\mathrm{min}}}\frac{\tilde{h}^{\mathrm{SR}}\tilde{h}^{\mathrm{SE}}}{S_n}df,
\end{equation}
where the mismatch is minimised over an arbitrary coalescence phase $\phi_c$ and time $t_c$ shift. $\tilde{h}$ denotes the Fourier transform $\tilde{h}(f)$ of $h(t)$. $S_n$ is the power spectral density of the detector, which we set to $S_n=1$ such that the mismatch is detector agnostic. We generate 100\,000 mismatches with $f_{\rm min} =5$\,Hz and $f_{\rm max} =2048$\,Hz using \textsc{IMRPhenomXAS\_NRTidalv3}, for randomly generated binaries with $m_{A,B}\in[1.0, 2.2]\,M_\odot$, $\Lambda_{A,B}\in [0,5000]$ and $\chi\in [0, 0.6]$. We keep all extrinsic parameters fixed. The mismatches are shown in Fig.~\ref{fig:mismatches} as a function of the effective spin $\chi_{\rm eff}  = (m_A\chi_A+m_B\chi_B)/(m_A+m_B)$ and effective tidal deformability $\tilde{\Lambda}$~\cite{Wade:2014vqa}. As expected, the mismatch is primarily sensitive to $\chi_{\rm eff}$, remaining below $10^{-4}$ for $\chi_{\rm eff}\lesssim 0.1$ but exceeding $10^{-2}$ for $\chi_{\rm eff}\gtrsim 0.3$. The largest mismatch achieved is $\mathcal{MM}\approx 0.25$, occurring at $\chi_{\rm eff}\approx 0.6$.

\begin{figure*}[tp]
    \centering
\captionsetup{justification=raggedright,singlelinecheck=false}
    \includegraphics[width=0.8\linewidth]{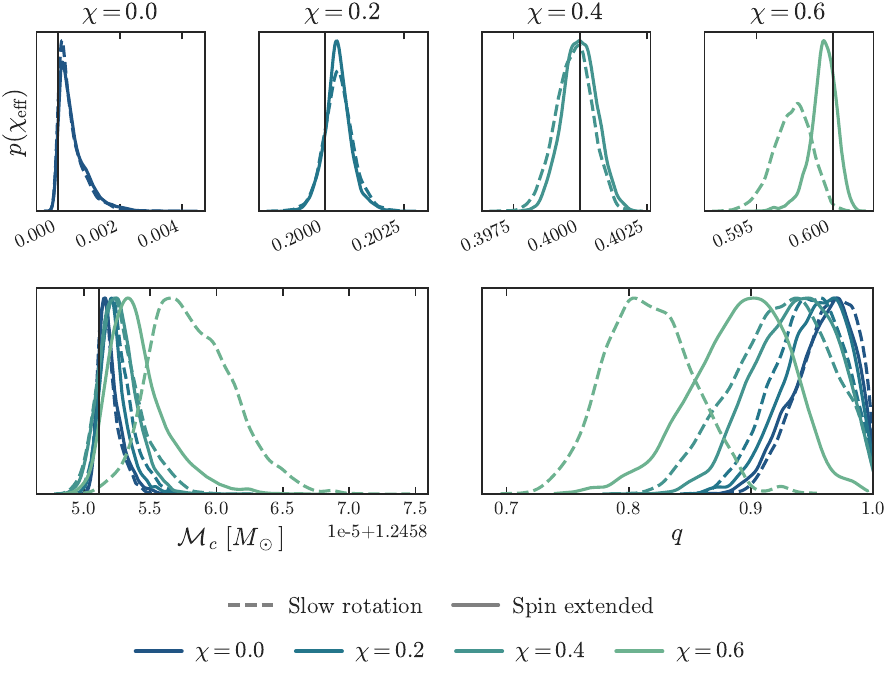}
    \caption{Marginalised posteriors for an equal-mass binary with $m_{A,B}=1.4\,M_\odot$ and equal spin $\chi_{A,B}\equiv\chi=[0.0,0.2,0.4,0.6]$ (colours) for the SLy EOS, injected with $C_Q$ as computed by \texttt{rns} and recovered assuming either the slow-rotation qUR (dashed) or the spin-extended qUR (solid). \textit{Upper row:} effective spin $\chi_{\rm eff}$, with one panel per injected spin. \textit{Lower row:} chirp mass $\mathcal{M}_c$ (left) and mass ratio $q$ (right), with all four injected spins overlaid. Injected values are marked in each panel (black solid vertical lines).}
    \label{fig:PE}
\end{figure*}

\subsection{Numerical-relativity comparison}

We additionally compare the impact of the spin-extended qUR with numerical-relativity data. It is well known that waveform models struggle to describe high-spinning BNS systems, with large discrepancies between models and numerical data. We compare waveforms generated using the slow-rotation qUR to those using the spin-extended qUR with model \textsc{IMRPhenomXAS\_NRTidalv3}. The numerical-relativity data~\cite{Kuan:2025bzu} uses the \texttt{BAM}~\cite{Bruegmann:2006ulg, Thierfelder:2011yi} code, and corresponds to two equal mass systems with equal spin $\chi = 0.5$, with EOSs SLy and H4 respectively. We use the highest resolution available. We align the waveforms by minimising the quantity
\begin{equation}
    \min_{t_0,\phi_0}\left[ \int^{t_2}_{t_1} \Big|\psi^{\mathrm{NR}}(t) - \psi^{\mathrm{model}}(t+t_0) - 2\psi_0\Big|^2 dt\right],
\end{equation}
across $t_1 = 500M$, $t_2 = 1500M$, to obtain time and phase shifts $t_0, \psi_0$. 
The numerical-relativity comparison is shown in Fig.~\ref{fig:NR}. For both SLy and H4 setups, the numerical-relativity data merges sooner than the waveform predicts, with a dephasing in the order of radians by merger. In comparison, the effect of the spin-extended qUR is of sub-radian order given the short length of the numerical-relativity waveform. In addition, the impact of the spin-extension in fact worsens the disagreement with numerical relativity - this is to be expected as we have shown that the slow-rotation qUR over-estimates $C_Q$ in this range of tidal deformabilities for large spins, thus resulting in a faster merger. Therefore although the spin-extended qUR reproduces more accurate values of $C_Q$, other spin induced mechanisms need to be investigated to improve the agreement with numerical-relativity simulations.

 \begin{figure*}
    \centering
\captionsetup{justification=raggedright,singlelinecheck=false}
    \includegraphics[width=0.8\linewidth]{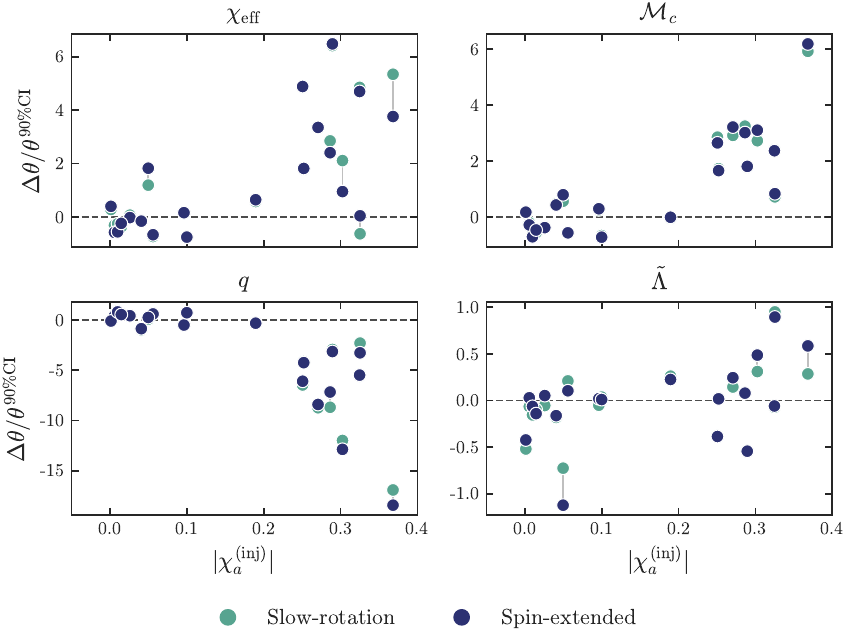}
    \caption{Biases $\Delta\theta/\theta^{90\%\mathrm{CI}} = (\theta^{\rm median}-\theta^{\rm true})/\theta^{90\%\mathrm{CI}}$ for the $20$ highest-SNR events from the population of $1000$ binaries with the EOS SLy, plotted against the injected spin asymmetry $|\chi_a^{(\rm inj)}|$, for analyses assuming the slow-rotation qUR (green) and the spin-extended qUR (blue), with grey lines connecting the two analyses of the same event. \textit{Upper row:} effective spin $\chi_{\rm eff}$ (left) and chirp mass $\mathcal{M}_c$ (right). \textit{Lower row:} mass ratio $q$ (left) and effective tidal deformability $\tilde{\Lambda}$ (right). The zero-bias line is marked in each panel (black dashed).}
    \label{fig:pop}
\end{figure*}

\section{Parameter Estimation}
\label{Sec:PE}
We next investigate the impact of the spin-extended qUR on observables from highly spinning BNS systems. This is achieved through Bayesian parameter estimation.
The posterior density distribution function of a set of model parameters $\boldsymbol{\theta}$ given the data $d$ is 
\begin{align}
    p(\boldsymbol{\theta}|d) &= \frac{ \mathcal{L}(\boldsymbol{\theta}) p(\boldsymbol{\theta})}{\mathcal{Z}_d},
\end{align}
where $\mathcal{L}(\boldsymbol{\theta})$ denotes the likelihood, $p(\boldsymbol{\theta})$ the prior, and $\mathcal{Z}_d$ the evidence or marginalised likelihood, defined as
\begin{equation}
    \mathcal{Z}_d = \int \mathcal{L}(\boldsymbol{\theta}) p (\boldsymbol{\theta}) d\boldsymbol{\theta}.
\end{equation}
The parameters $\boldsymbol{\theta}$ that we consider are the intrinsic binary parameters such as the component masses $m_{A,B}$, the dimensionless aligned spin of the components $\chi_{A,B}$, and the quadrupolar tidal deformabilites $\Lambda_{A,B}$, and extrinsic parameters, i.e., the sky location, inclination, distance, polarisation, coalescence time and phase. We obtain one-dimensional and two-dimensional posteriors by marginalising over other parameters. We model the GW signal with the \textsc{IMRPhenomXAS\_NRTidalv3} waveform model~\cite{Pratten:2020fqn, Abac:2023ujg}, which accounts for unequal-mass, aligned-spin BNS systems. Parameter estimation is performed with the \textsc{Bilby} inference library~\cite{Ashton:2018jfp} in conjunction with the nested sampler \textsc{Dynesty}~\cite{Speagle:2019ivv}. We choose a configuration of two 15\,km L-shaped Einstein Telescope detectors placed in Limburg and Sardinia respectively with power spectral densities as in Ref.~\cite{et_psd}. We use a minimum frequency of 5\,Hz, a maximum of 2048\,Hz, and a sampling rate of 4096\,Hz in all analyses. For parameter estimation acceleration, we apply a multibanded likelihood~\cite{Vinciguerra:2017ngf,Morisaki:2021ngj} and marginalise over the luminosity distance $D_L$. We sample in the chirp mass $\mathcal{M}_c$, mass ratio $q$, aligned component spins $\chi_A$ and $\chi_B$, tidal parameters $\tilde{\Lambda}$ and $\delta\tilde{\Lambda}$~\cite{Wade:2014vqa}, source-frame luminosity distance $D_L$, and sky-location and orientation angles, which are assumed to be isotropically distributed. Priors are uniform in component masses, aligned spins, $\tilde{\Lambda}$, $\delta\tilde{\Lambda}$, and $D_L$ in comoving volume and source frame time, with ranges $\mathcal{M}_c \in [0.7, 2.0]\,M_\odot$, $q \in [0.4, 1]$, $\chi_{A,B} \in [-0.65, 0.65]$, $D_L \in [0, 1200] \,\mathrm{Mpc}$, $\tilde{\Lambda} \in [0, 5000]$, and $\delta\tilde{\Lambda} \in [-5000, 5000]$. We run parameter estimation with zero noise, such that any observed biases arise solely from the difference in qURs and not from random noise fluctuations. 

We first select a canonical equal-mass $m_{A,B}=1.4M_{\odot}$ system, with EOS SLy corresponding to $\Lambda_{A,B}=302$, and set $D_L=100$\ Mpc. We simulate the injected waveforms such that $C_Q$ is set from the calculated \texttt{rns} value (with the exception of $\chi_{A,B}=0$ for which the SIQM vanishes identically). We then perform parameter estimation twice, once assuming the slow-rotation qUR as is currently implemented, and once assuming the spin-extended qUR. We do this for systems of equal-spin $\chi_{A,B}\equiv\chi=[0, 0.2, 0.4, 0.6]$.

Figure~\ref{fig:PE} shows the posteriors for $\chi_{\rm eff}$, $\mathcal{M}_c$ and $q$. We omit $\tilde{\Lambda}$, which is consistently recovered for all systems. At $\chi=0$ the slow-rotation and spin-extended qURs give identical posteriors, both recovering the injected values, as expected. Biases in all three parameters then grow with spin, a consequence of the well-known mass-spin degeneracy. At the moderate spins $\chi=0.2$ and $0.4$, the two relations recover $\chi_{\rm eff}$ and $\mathcal{M}_c$ comparably well and differ only in $q$, where the slow-rotation posteriors are biased slightly further from the equal-mass boundary. The difference becomes substantial at $\chi=0.6$, where the slow-rotation qUR biases $\chi_{\rm eff}$ low, $\mathcal{M}_c$ high and $q$ low, placing the injected values outside the $90\%$ credible intervals (CIs). The spin-extended qUR, by contrast, still recovers the injected $\chi_{\rm eff}$ and shows less bias in the masses. For this setup, the spin-extended qUR reduces biases across the intrinsic parameters.

Next we study posteriors from a moderate-to-highly spinning BNS population. We simulate 1000 binaries with component masses drawn from the \textsc{FLAT\_Q} model distribution~\cite{Landry:2021hvl} in interval $m\in[1,M_{\rm TOV}] \, M_{\odot}$ with $M_{\rm TOV}=2.04\, M_{\odot}$, aligned spins drawn uniformly and restricted such that $|\chi|\in[0.15, 0.6]$, and luminosity distance drawn uniformly in comoving volume $D_L\in[10,1000]$\,Mpc, and isotropically distributed sky location. We again use the SLy EOS to calculate the $\Lambda$ and $C_Q$ using \texttt{rns}. From this population we then select the 20 events with the largest SNRs (ranging from 177 to 306) to perform parameter estimation. We quantify the bias of a parameter $\theta$ with 
\begin{equation}
\Delta\theta/\theta^{\rm 90\% CI}=(\theta^{\rm median}-\theta^{\rm true})/\theta^{\rm 90\% CI},
\end{equation}
where $\theta^{\rm 90\% CI}$ is the width of the 90\% confidence interval (CI). Normalising in this way measures the systematic offset against the statistical uncertainty, which at these SNRs is small: the median absolute biases $|\Delta\theta|$ are $7.1\times10^{-6}\,M_\odot$ in $\mathcal{M}_c$, $3.8\times10^{-3}$ in $\chi_{\rm eff}$, $0.065$ in $q$ and $5.3$ in $\tilde{\Lambda}$.

Figure \ref{fig:pop} shows the biases in $\chi_{\rm eff}$, $\mathcal{M}_c$, $q$ and $\tilde{\Lambda}$. As in the single-system case, bias in $\chi_{\rm eff}$, $\mathcal{M}_c$ and $q$ grows with the spin content of the system. However unlike Fig. 8, where $\chi_A=\chi_B$ by construction, the population's component spins are drawn independently, and we find $\chi_{\rm eff}$ is in fact a poor predictor of bias here where negative spins are permitted. Instead, the bias correlates more with the spin asymmetry $\chi_a=(\chi_A-\chi_B)/2$. We observe here that the improvement of the bias is unreliable, with roughly half systems showing improvement in the bias for a given parameter when using the spin-extended qUR. For systems with large spin asymmetry $|\chi_a|>0.3$, we find that the spin-extended qUR reduces the bias in $\chi_{\rm eff}$ in all cases. However, this appears to come at the expense of increased biases in the mass parameters $\mathcal{M}_c$ and $q$. This demonstrates that the mass-spin degeneracy dominates at the population level.
\section{Conclusion}
\label{Sec:conclusion}
We have presented a qUR between the tidal deformability and the SIQM that is extended for rapidly rotating NSs. Calibrated against $C_Q$ computed with \texttt{rns} for 128 cold hadronic, hyperonic and
hybrid tabulated EOSs over $\chi\in[0.15,0.6]$, this qUR is a drop-in replacement for the one currently used in
waveform models: it introduces no additional EOS-dependent parameters, and recovers the slow-rotation qUR in the non-spinning limit. This spin-extended qUR reduces the median absolute percentage error on a validation set from $4.88\%$ to $1.66\%$. 

The improvement in $C_Q$ translates into a modest but non-negligible
change in the waveform. The phase accumulated from 5\,Hz differs by up to $\sim2$~rad for an equal-mass system with $\Lambda_{A,B}=500$ and $\chi_{A,B}=0.4$, with mismatches between waveforms reaching $\approx0.25$ at the highest spins. 

The improvement in $C_Q$ does not translate uniformly into reduced bias in parameter estimation. For a single equal-spin SLy binary the spin-extended relation reduces bias across the intrinsic parameters, however, across a population with independently drawn components, mass-spin degeneracies dominate during parameter estimation. Systems with $|\chi_a|>0.3$ show a reduced bias in
$\chi_{\rm eff}$ in all cases, but at the cost of increased bias in
$\mathcal{M}_c$ and $q$. 

Despite the moderate improvement in parameter estimation and gravitational-wave dephasing, the spin-extended qUR presented here provides a more accurate description of the SIQM than the slow-rotation relation over the full range of spins considered. Its effect on inference is currently masked by the mass-spin degeneracy, but the relation removes a known and quantifiable inaccuracy from the waveform at no cost in model dimensionality, and is directly applicable in existing approximants. Possible future work includes extending to the spin-induced octupolar tidal deformability $C_{\rm Oc}$ for which the qUR also uses the slow-rotation approximation. Additionally, fitting $C_Q$ against the rotational tidal Love numbers~\cite{Pani:2015nua, Abdelsalhin:2018reg} rather than the non-rotating $\Lambda$ may improve universality, since both parameters would then describe the same rotating configuration, removing the inconsistency responsible for the turning point in the spin-extended fit presented here. These are, however, known only to linear order in spin and thus may themselves break down at the large spins considered here, and are not commonly included in current waveform models.

In addition our results here highlight that future research efforts should be directed towards including additional spin-matter mechanisms to bring waveform models closer to numerical-relativity data in the high-spin regime. These mechanisms may include spin-tidal couplings~\cite{Pani:2015nua, Abdelsalhin:2018reg, JimenezForteza:2018rwr} and spin-shifted dynamical tidal effects~\cite{Steinhoff:2021dsn,Kuan:2022etu}. 
\section*{Acknowledgments}
The authors thank Ivan Markin for supplying numerical-relativity data. In addition, the authors thank Guilherme Grams, Anna Puecher, Adrian Abac, and J\'{a}nos Tak\'{a}tsy for useful discussions and comments.

T.D. acknowledges funding from the EU Horizon under ERC Starting Grant, no. SMArt-101076369. Views and opinions expressed are those of the authors only and do not necessarily reflect those of the European Union or the European Research Council. Neither the European Union nor the granting authority can be held responsible for them. 

Computations were performed on the DFG-funded research cluster Jarvis at the University of Potsdam (INST 336/173-1; project number: 502227537).

This research has made use of data or software obtained from the Gravitational Wave Open Science Center (gwosc.org), a service of the LIGO Scientific Collaboration, the Virgo Collaboration, and KAGRA. This material is based upon work supported by NSF's LIGO Laboratory which is a major facility fully funded by the National Science Foundation, as well as the Science and Technology Facilities Council (STFC) of the United Kingdom, the Max-Planck-Society (MPS), and the State of Niedersachsen/Germany for support of the construction of Advanced LIGO and construction and operation of the GEO600 detector. Additional support for Advanced LIGO was provided by the Australian Research Council. Virgo is funded, through the European Gravitational Observatory (EGO), by the French Centre National de Recherche Scientifique (CNRS), the Italian Istituto Nazionale di Fisica Nucleare (INFN) and the Dutch Nikhef, with contributions by institutions from Belgium, Germany, Greece, Hungary, Ireland, Japan, Monaco, Poland, Portugal, Spain. KAGRA is supported by Ministry of Education, Culture, Sports, Science and Technology (MEXT), Japan Society for the Promotion of Science (JSPS) in Japan; National Research Foundation (NRF) and Ministry of Science and ICT (MSIT) in Korea; Academia Sinica (AS) and National Science and Technology Council (NSTC) in Taiwan.
\appendix
\section{Equations of State}
\label{Appdx:EOS}
\onecolumngrid
\begingroup
\footnotesize
\setlength{\tabcolsep}{4pt}
\renewcommand{\arraystretch}{0.9}
\begin{longtable}{l l c c c l}
\label{tab:eos_summary} \\
\toprule
EOS & Particles & $M_{\rm TOV}\,[M_\odot]$ & $R_{1.4}\,[\rm km]$ & $\Lambda_{1.4}$ & Ref. \\
\midrule
\endfirsthead
\multicolumn{6}{c}{{\tablename\ \thetable{} -- continued from previous page}} \\
\toprule
EOS & Particles & $M_{\rm TOV}\,[M_\odot]$ & $R_{1.4}\,[\rm km]$ & $\Lambda_{1.4}$ & Ref. \\
\midrule
\endhead
\midrule
\multicolumn{6}{r}{{Continued on next page}} \\
\endfoot
\bottomrule
\endlastfoot
DDHdelta & $npe$ & 2.15 & 12.69 & 584 & \cite{Gaitanos:2003zg, Grill:2014aea, Douchin:2001sv} \\
GM(GM1) & $npe$ & 2.39 & 14.24 & 982 & \cite{Glendenning:1991es, Douchin:2001sv} \\
GM1Y4 & $npeB_s$ & 1.79 & 13.82 & 979 & \cite{Glendenning:1991es, Oertel:2014qza, Douchin:2001sv} \\
GM1Y5 & $npeB_s$ & 2.12 & 13.82 & 979 & \cite{Glendenning:1991es, Oertel:2014qza, Douchin:2001sv} \\
GM1Y6 & $npeB_s$ & 2.30 & 13.82 & 977 & \cite{Glendenning:1991es, Oertel:2014qza, Douchin:2001sv} \\
DDHdeltaY4 & $npeB_s$ & 2.04 & 12.69 & 585 & \cite{Gaitanos:2003zg, Oertel:2014qza, Grill:2014aea, Douchin:2001sv} \\
APR(APR) & $npe\mu$ & 2.10 & 11.38 & 252 & \cite{Akmal:1998cf, Baym:1971pw, Douchin:2001sv} \\
SLY230a & $npe\mu$ & 2.10 & 11.89 & 332 & \cite{Danielewicz:2008cm, Gulminelli:2015csa, Chabanat:1997un} \\
SLY2 & $npe\mu$ & 2.06 & 11.84 & 314 & \cite{Danielewicz:2008cm, Chabanat:1995tea, Gulminelli:2015csa} \\
SLY9 & $npe\mu$ & 2.16 & 12.53 & 455 & \cite{Danielewicz:2008cm, Chabanat:1995tea, Gulminelli:2015csa} \\
SkI2 & $npe\mu$ & 2.17 & 13.55 & 779 & \cite{Reinhard:1995zz, Danielewicz:2008cm, Gulminelli:2015csa} \\
SkI3 & $npe\mu$ & 2.24 & 13.62 & 795 & \cite{Reinhard:1995zz, Danielewicz:2008cm, Gulminelli:2015csa} \\
SkI4 & $npe\mu$ & 2.17 & 12.43 & 473 & \cite{Reinhard:1995zz, Danielewicz:2008cm, Gulminelli:2015csa} \\
SkI5 & $npe\mu$ & 2.24 & 14.15 & 1018 & \cite{Reinhard:1995zz, Danielewicz:2008cm, Gulminelli:2015csa} \\
SkI6 & $npe\mu$ & 2.19 & 12.55 & 495 & \cite{Reinhard:1995zz, Danielewicz:2008cm, Gulminelli:2015csa} \\
KDE0v1 & $npe\mu$ & 1.97 & 11.68 & 271 & \cite{Danielewicz:2008cm, Gulminelli:2015csa, Agrawal:2005ix} \\
KDE0v & $npe\mu$ & 1.96 & 11.47 & 245 & \cite{Danielewicz:2008cm, Gulminelli:2015csa, Agrawal:2005ix} \\
SK255 & $npe\mu$ & 2.15 & 13.21 & 596 & \cite{Danielewicz:2008cm, Agrawal:2003xb, Gulminelli:2015csa} \\
SK272 & $npe\mu$ & 2.23 & 13.38 & 653 & \cite{Danielewicz:2008cm, Agrawal:2003xb, Gulminelli:2015csa} \\
SKa & $npe\mu$ & 2.21 & 12.98 & 574 & \cite{Kohler:1976fgx, Danielewicz:2008cm, Gulminelli:2015csa} \\
SKb & $npe\mu$ & 2.19 & 12.26 & 467 & \cite{Kohler:1976fgx, Danielewicz:2008cm, Gulminelli:2015csa} \\
SkMp & $npe\mu$ & 2.11 & 12.56 & 486 & \cite{Danielewicz:2008cm, Bennour:1989zz, Gulminelli:2015csa} \\
SkOp & $npe$ & 1.98 & 12.18 & 368 & \cite{Reinhard:1995zz, Danielewicz:2008cm, Gulminelli:2015csa} \\
Rs & $npe\mu $ & 2.12 & 12.99 & 599 & \cite{Danielewicz:2008cm, Friedrich:1986zza, Gulminelli:2015csa},  \\
chiral with crust & $npe\mu $ & 2.08 & 12.36 & 391 & \cite{Douchin:2001sv, Bombaci:2018ksa} \\
SLY4 & $npe\mu N$ & 2.04 & 11.76 & 302 & \cite{Danielewicz:2008cm, Gulminelli:2015csa, Chabanat:1997un} \\
CMF with crust & $npe\mu B_s$ & 2.07 & 13.63 & 869 & \cite{Dexheimer:2008ax, Schurhoff:2010ph, Dexheimer:2015qha, Dexheimer:2017nse} \\
QHC18 & $npeNq$ & 2.04 & 11.54 & 260 & \cite{Akmal:1998cf, Togashi:2017mjp, Baym:2017whm} \\
QHC19-B & $npeNq$ & 2.07 & 11.62 & 312 & \cite{Togashi:2017mjp, Baym:2017whm, Baym:2019iky} \\
DD2-FRG (2+1 flavors) & $npeNqs$ & 1.89 & 13.27 & 707 & \cite{Hempel:2009mc, Typel:2009sy, Otto:2019zjy, Otto:2020hoz} \\
DD2-FRG (2 flavors) & $npeNq$ & 2.07 & 13.28 & 707 & \cite{Hempel:2009mc, Typel:2009sy, Otto:2019zjy, Otto:2020hoz} \\
QHC19-A & $npeNq$ & 1.93 & 11.58 & 303 & \cite{Togashi:2017mjp, Baym:2017whm, Baym:2019iky} \\
QHC19-C & $npeNq$ & 2.18 & 11.64 & 315 & \cite{Togashi:2017mjp, Baym:2017whm, Baym:2019iky} \\
QHC19-D & $npeNq$ & 2.28 & 11.64 & 315 & \cite{Togashi:2017mjp, Baym:2017whm, Baym:2019iky} \\
CMF-1 with crust & $npeNB_s$ & 2.07 & 13.63 & 868 & \cite{Bennour:1989zz, Gulminelli:2015csa, Dexheimer:2008ax, Dexheimer:2017nse,Dexheimer:2009hi, Dexheimer:2020rlp} \\
CMF-2 with crust & $npeN$ & 2.13 & 13.76 & 933 & \cite{Bennour:1989zz, Gulminelli:2015csa, Dexheimer:2008ax, Dexheimer:2017nse,Dexheimer:2009hi, Dexheimer:2020rlp} \\
CMF-3 with crust & $npeNB_s$ & 2.01 & 13.23 & 708 & \cite{Bennour:1989zz, Gulminelli:2015csa, Dexheimer:2008ax, Dexheimer:2017nse,Dexheimer:2009hi, Dexheimer:2020rlp} \\
CMF-4 with crust & $npeN$ & 2.05 & 13.33 & 747 & \cite{Bennour:1989zz, Gulminelli:2015csa, Dexheimer:2008ax, Dexheimer:2017nse,Dexheimer:2009hi, Dexheimer:2020rlp} \\
CMF-5 with crust & $npeNB_s$ & 2.08 & 13.27 & 729 & \cite{Bennour:1989zz, Gulminelli:2015csa, Dexheimer:2008ax, Dexheimer:2017nse,Dexheimer:2009hi, Dexheimer:2020rlp} \\
CMF-6 with crust & $npeN$ & 2.11 & 13.36 & 766 & \cite{Bennour:1989zz, Gulminelli:2015csa, Dexheimer:2008ax, Dexheimer:2017nse,Dexheimer:2009hi, Dexheimer:2020rlp} \\
CMF-7 with crust & $npeNB_s \Delta$ & 2.08 & 13.28 & 729 & \cite{Bennour:1989zz, Gulminelli:2015csa, Dexheimer:2008ax, Dexheimer:2017nse,Dexheimer:2009hi, Dexheimer:2020rlp} \\
CMF-8 with crust & $npeN\Delta$ & 2.09 & 13.36 & 766 & \cite{Bennour:1989zz, Gulminelli:2015csa, Dexheimer:2008ax, Dexheimer:2017nse,Dexheimer:2009hi, Dexheimer:2020rlp} \\
DD2-FRG w/ vector interactions (2 flavors) & $npeNq$ & 2.15 & 13.27 & 707 & \cite{Hempel:2009mc, Typel:2009sy, Otto:2019zjy, Otto:2020hoz} \\
DD2-FRG w/ vector interactions (2+1 flavors) & $npeNqs$ & 1.84 & 13.27 & 706 & \cite{Hempel:2009mc, Typel:2009sy, Otto:2019zjy, Otto:2020hoz} \\
VQCD(APR), intermediate & $npeNq$ & 2.14 & 12.44 & 497 & \cite{Akmal:1998cf, Jokela:2018ers, Ishii:2019gta, Ecker:2019xrw, Jokela:2020piw} \\
VQCD(APR), stiff & $npNq$ & 2.33 & 12.60 & 545 & \cite{Akmal:1998cf, Jokela:2018ers, Ishii:2019gta, Ecker:2019xrw, Jokela:2020piw} \\
VQCD(APR), soft & $npeNq$ & 2.01 & 12.38 & 475 & \cite{Akmal:1998cf, Jokela:2018ers, Ishii:2019gta, Ecker:2019xrw, Jokela:2020piw} \\
FSU2H unified inner crust-core & $npe\mu N$ & 2.38 & 13.34 & 739 & \cite{Grill:2014aea, Negreiros:2018cho, Providencia:2018ywl, Pearson:2018tkr} \\
FSU2R unified inner crust-core & $npe\mu N$ & 2.05 & 12.99 & 599 & \cite{Grill:2014aea, Negreiros:2018cho, Providencia:2018ywl, Pearson:2018tkr} \\
FSU2 unified inner crust-core & $npe\mu N$ & 2.07 & 13.93 & 872 & \cite{Grill:2014aea, Negreiros:2018cho, Providencia:2018ywl, Pearson:2018tkr} \\
SLy5 & $npe\mu N$ & 2.10 & 11.83 & 316 & \cite{Grams:2022lci} \\
DD2 unified inner crust-core & $npe\mu N$ & 2.42 & 13.24 & 695 & \cite{Typel:2009sy, Grill:2014aea, Pearson:2018tkr} \\
DDME2 unified inner crust-core & $npe\mu N$ & 2.48 & 13.28 & 715 & \cite{Grill:2014aea, Lalazissis:2005de, Pearson:2018tkr} \\
TW unified inner crust-core & $npe\mu N$ & 2.08 & 12.37 & 411 & \cite{Grill:2014aea, Typel:1999yq, Pearson:2018tkr} \\
NL3wrL55 unified inner crust-core & $npe\mu N$ & 2.76 & 13.79 & 934 & \cite{Grill:2014aea, Horowitz:2001ya, Horowitz:2000xj} \\
TM1e unified inner crust-core & $npe\mu N$ & 2.12 & 13.21 & 655 & \cite{Grill:2014aea, Shen:2020sec, Pearson:2018tkr} \\
LNS5 & $npe$ & 1.97 & 11.66 & 294 & \cite{Grill:2014aea, Shen:2020sec, Pearson:2018tkr} \\
BSK16 & $npe\mu N$ & 1.85 & -- & -- & \cite{Grams:2022lci} \\
BSK14 & $npe\mu N$ & 1.92 & 11.21 & 217 & \cite{Grams:2022lci} \\
RATP & $npe\mu N$ & 1.75 & -- & -- & \cite{Grams:2022lci} \\
SGII & $npe$ & 1.41 & 10.15 & 108 & \cite{Grams:2022lci} \\
F0 & $npe\mu N$ & 2.07 & 11.72 & 302 & \cite{Grams:2022lci} \\
H1 & $npe\mu N$ & 2.29 & 11.53 & 297 & \cite{Grams:2022lci} \\
H2 & $npe\mu N$ & 2.28 & 11.72 & 328 & \cite{Grams:2022lci} \\
H3 & $npe\mu N$ & 2.28 & 12.07 & 389 & \cite{Grams:2022lci} \\
H4 & $npe\mu N$ & 2.30 & 11.88 & 354 & \cite{Grams:2022lci} \\
H5 & $npe\mu N$ & 2.36 & 12.19 & 403 & \cite{Grams:2022lci} \\
H7 & $npe\mu N$ & 2.48 & 12.85 & 543 & \cite{Grams:2022lci} \\
DHSL59 & $npe\mu N$ & 2.28 & 12.56 & 467 & \cite{Grams:2022lci} \\
DHSL69 & $npe\mu N$ & 2.32 & 12.61 & 475 & \cite{Grams:2022lci} \\
QHC21\_AT & $npeNq$ & 2.13 & 11.88 & 375 & \cite{Togashi:2017mjp, Kojo:2021wax} \\
QHC21\_BT & $npeNq$ & 2.20 & 11.91 & 382 & \cite{Togashi:2017mjp, Kojo:2021wax} \\
QHC21\_CT & $npeNq$ & 2.26 & 11.80 & 356 & \cite{Togashi:2017mjp, Kojo:2021wax} \\
QHC21\_DT & $npeNq$ & 2.31 & 11.81 & 358 & \cite{Togashi:2017mjp, Kojo:2021wax} \\
QHC21\_A & $npeNq$ & 2.19 & 12.42 & 461 & \cite{Togashi:2017mjp, Kojo:2021wax, Drischler:2020fvz} \\
QHC21\_B & $npeNq$ & 2.25 & 12.42 & 462 & \cite{Togashi:2017mjp, Kojo:2021wax, Drischler:2020fvz} \\
QHC21\_C & $npeNq$ & 2.31 & 12.42 & 460 & \cite{Togashi:2017mjp, Kojo:2021wax, Drischler:2020fvz} \\
QHC21\_D & $npeNq$ & 2.36 & 12.41 & 459 & \cite{Togashi:2017mjp, Kojo:2021wax, Drischler:2020fvz} \\
DD2YDelta 1.1-1.1 & $npeNB_s\Delta$ & 2.04 & 13.03 & 584 & \cite{Typel:2009sy, Raduta:2020fdn, Raduta:2022elz, Vinas:2021vmv} \\
DD2YDelta 1.2-1.3 & $npNB_s\Delta$ & 2.03 & 13.29 & 691 & \cite{Typel:2009sy, Raduta:2020fdn, Raduta:2022elz, Vinas:2021vmv} \\
DD2YDelta 1.2-1.1 & $npNB_s\Delta$ & 2.05 & 12.35 & 392 & \cite{Typel:2009sy, Raduta:2020fdn, Raduta:2022elz, Vinas:2021vmv} \\
BSK24 & $npe\mu N$ & 2.28 & 12.61 & 520 & \cite{Audi:2017asy, Allard:2021rrt, Pearson:2020bxz, Pearson:2022vep, Pearson:2018tkr, Goriely:2013nxa, Perot:2019gwl, Xu:2012uw, Welker:2017eja} \\
D1MStar & $npe\mu N$ & 2.01 & 11.75 & 318 & \cite{Vinas:2021vmv, Mondal:2020cgu, Gonzalez-Boquera:2017rzy} \\
D1M & $npe\mu N$ & 1.71 & 10.24 & 123 & \cite{Vinas:2021vmv, Mondal:2020cgu, Gonzalez-Boquera:2017rzy} \\
BSK22 & $npe\mu N$ & 2.27 & 13.08 & 634 & \cite{Audi:2017asy, Allard:2021rrt, Pearson:2020bxz, Pearson:2022vep, Pearson:2018tkr, Goriely:2013nxa, Perot:2019gwl, Xu:2012uw, Welker:2017eja} \\
BSK25 & $npe\mu N$ & 2.23 & 12.41 & 480 & \cite{Audi:2017asy, Allard:2021rrt, Pearson:2020bxz, Pearson:2022vep, Pearson:2018tkr, Goriely:2013nxa, Perot:2019gwl, Xu:2012uw, Welker:2017eja} \\
BSK26 & $npe\mu N$ & 2.17 & 11.81 & 328 & \cite{Audi:2017asy, Allard:2021rrt, Pearson:2020bxz, Pearson:2022vep, Pearson:2018tkr, Goriely:2013nxa, Perot:2019gwl, Xu:2012uw, Welker:2017eja} \\
CMF-1 Hybrid with crust & $npeNB_sq$ & 1.97 & 13.63 & 868 & \cite{Dexheimer:2008ax, Dexheimer:2017nse, Dexheimer:2009hi, Dexheimer:2020rlp, Clevinger:2022xzl, Dexheimer:2018dhb} \\
CMF-2 Hybrid with crust & $npeNq$ & 1.96 & 13.76 & 932 & \cite{Dexheimer:2008ax, Dexheimer:2017nse, Dexheimer:2009hi, Dexheimer:2020rlp, Clevinger:2022xzl, Dexheimer:2018dhb} \\
CMF-3 Hybrid with crust & $npeNB_sq$ & 1.99 & 13.23 & 714 & \cite{Dexheimer:2008ax, Dexheimer:2017nse, Dexheimer:2009hi, Dexheimer:2020rlp, Clevinger:2022xzl, Dexheimer:2018dhb} \\
CMF-4 Hybrid with crust & $npeNq$ & 1.98 & 13.33 & 753 & \cite{Dexheimer:2008ax, Dexheimer:2017nse, Dexheimer:2009hi, Dexheimer:2020rlp, Clevinger:2022xzl, Dexheimer:2018dhb} \\
CMF-5 Hybrid with crust & $npeNB_sq$ & 2.02 & 13.28 & 734 & \cite{Dexheimer:2008ax, Dexheimer:2017nse, Dexheimer:2009hi, Dexheimer:2020rlp, Clevinger:2022xzl, Dexheimer:2018dhb} \\
CMF-6 Hybrid with crust & $npeNq$ & 2.02 & 13.37 & 772 & \cite{Dexheimer:2008ax, Dexheimer:2017nse, Dexheimer:2009hi, Dexheimer:2020rlp, Clevinger:2022xzl, Dexheimer:2018dhb} \\
CMF-7 Hybrid with crust & $npeNB_sq\Delta$ & 2.02 & 13.28 & 735 & \cite{Dexheimer:2008ax, Dexheimer:2017nse, Dexheimer:2009hi, Dexheimer:2020rlp, Clevinger:2022xzl, Dexheimer:2018dhb} \\
CMF-8 Hybrid with crust & $npeN\Delta q$ & 2.02 & 13.37 & 772 & \cite{Dexheimer:2008ax, Dexheimer:2017nse, Dexheimer:2009hi, Dexheimer:2020rlp, Clevinger:2022xzl, Dexheimer:2018dhb} \\
QMC-RMF1 & $npeN$ & 1.95 & 11.88 & 313 & \cite{Grill:2014aea, Baym:1971pw, Alford:2022bpp} \\
QMC-RMF2 & $npeN$ & 2.04 & 12.05 & 356 & \cite{Grill:2014aea, Baym:1971pw, Alford:2022bpp} \\
QMC-RMF3 & $npeN$ & 2.15 & 12.28 & 388 & \cite{Grill:2014aea, Baym:1971pw, Alford:2022bpp} \\
QMC-RMF4 & $npeN$ & 2.21 & 12.38 & 404 & \cite{Grill:2014aea, Baym:1971pw, Alford:2022bpp} \\
DDME2 & $npe\mu N$ & 2.49 & 13.24 & 718 & \cite{Lalazissis:2005de, Xia:2022dvw, Xia:2022pja, Niu:2025tvd} \\
DD-LZ1 & $npe\mu N$ & 2.56 & 13.19 & 722 & \cite{Xia:2022dvw, Xia:2022pja, Wei:2020kfb, Niu:2025tvd} \\
DDME-X & $npe\mu N$ & 2.56 & 13.41 & 787 & \cite{Xia:2022dvw, Xia:2022pja, Taninah:2019cku, Niu:2025tvd} \\
XMLSLZ(GM1) & $npe\mu N$ & 2.37 & 13.79 & 921 & \cite{Glendenning:1991es, Xia:2022dvw, Xia:2022pja, Niu:2025tvd} \\
MTVTC & $npe\mu N$ & 2.02 & 13.13 & 631 & \cite{Xia:2022dvw, Xia:2022pja, Niu:2025tvd, Maruyama:2005vb} \\
NL3 & $npe\mu N$ & 2.78 & 14.64 & 1301 & \cite{Lalazissis:2005de, Xia:2022dvw, Xia:2022pja, Niu:2025tvd} \\
PK1 & $npe\mu N$ & 2.31 & 14.41 & 1118 & \cite{Xia:2022dvw, Xia:2022pja, Niu:2025tvd, Long:2003dn} \\
PKDD & $npe\mu N$ & 2.33 & 13.67 & 776 & \cite{Xia:2022dvw, Xia:2022pja, Niu:2025tvd, Long:2003dn} \\
TM1 & $npe\mu N$ & 2.18 & 14.31 & 1065 & \cite{Sugahara:1993wz, Xia:2022dvw, Xia:2022pja} \\
TW99 & $npe\mu N$ & 2.08 & 12.30 & 411 & \cite{Typel:1999yq, Xia:2022dvw, Xia:2022pja, Niu:2025tvd} \\
PCSB0 & $npeN$ & 2.53 & 13.35 & 715 & \cite{Hempel:2009mc, Hornick:2018kfi, Pradhan:2022txg},  \\
PCSB1 & $npe\mu N$ & 2.19 & 13.13 & 630 & \cite{Hempel:2009mc, Hornick:2018kfi, Pradhan:2022txg} \\
PCSB2 & $npe\mu $ & 2.02 & 12.93 & 559 & \cite{Hempel:2009mc, Hornick:2018kfi, Pradhan:2022txg} \\
GRDF2-DD2 & $npe\mu N$ & 2.42 & 13.21 & 690 & \cite{Hempel:2009mc, Fattoyev:2010mx, Alford:2022bpp, Alford:2023rgp} \\
chiral with unified crust & $npe\mu N$ & 2.08 & 12.32 & 392 & \cite{Bombaci:2018ksa, Carreau:2019zdy} \\
GDFM-I & $npe\mu N$ & 2.31 & 12.87 & 536 & \cite{Carreau:2019zdy, Gogelein:2007qa, Char:2023fue} \\
GDFM-II & $npe\mu N$ & 2.31 & 13.88 & 927 & \cite{Carreau:2019zdy, Gogelein:2007qa, Char:2023fue} \\
M1 unified & $npe\mu N$ & 2.54 & 12.84 & 541 & \cite{Pearson:2018tkr, Gogelein:2007qa, Scurto:2024ekq} \\
M2 unified & $npe\mu N$ & 2.42 & 12.66 & 525 & \cite{Pearson:2018tkr, Gogelein:2007qa, Scurto:2024ekq} \\
M3 unified & $npe\mu N$ & 2.68 & 12.70 & 523 & \cite{Pearson:2018tkr, Gogelein:2007qa, Scurto:2024ekq} \\
M4 unified & $npe\mu N$ & 2.34 & 12.32 & 438 & \cite{Pearson:2018tkr, Gogelein:2007qa, Scurto:2024ekq} \\
M5 unified & $npe\mu N$ & 2.71 & 13.45 & 770 & \cite{Pearson:2018tkr, Gogelein:2007qa, Scurto:2024ekq} \\
QMC-RMF1 unified crust & $npeN$ & 1.95 & 11.79 & 315 & \cite{Alford:2022bpp, Davis:2024nda, Davis:2025nwz} \\
DDFGOS(APR) & $npeN$ & 2.10 & 11.53 & 259 & \cite{Akmal:1998cf, Davis:2024nda, Constantinou:2014hha, Davis:2025nwz} \\
APR unified crust & $npe$ & 2.12 & 11.41 & 252 & \cite{Akmal:1998cf, Davis:2024nda, Davis:2025nwz, Haensel:2007yy} \\
DDME-X1 & $npe\mu N$ & 2.62 & 13.59 & 864 & \cite{Xia:2022dvw, Xia:2022pja, Niu:2025tvd, Taninah:2019cku} \\
DDME-X2 & $npe\mu N
$ & 2.51 & 13.82 & 900 & \cite{Xia:2022dvw, Xia:2022pja, Niu:2025tvd, Taninah:2019cku} \\
DDME-Y & $npe\mu N$ & 2.55 & 13.36 & 772 & \cite{Xia:2022dvw, Xia:2022pja, Niu:2025tvd, Taninah:2019cku} \\
PK1r & $npe\mu N$ & 2.32 & 14.38 & 1109 & \cite{Xia:2022dvw, Xia:2022pja, Niu:2025tvd, Long:2003dn} \\
\caption{Summary of the 128 cold, tabulated EOSs used in the fitting of
the spin-extended qUR presented in this work. All EOS
tables were obtained from the CompOSE database~\cite{Typel:2013rza,CompOSECoreTeam:2022ddl, CompOSE}. We list the particles included at $\beta$-equilibrium: neutrons $n$,
protons $p$, electrons $e$, muons $\mu$, nuclei $N$, hyperons $B_s$,
$\Delta(1232)$ resonances $\Delta$, deconfined quarks $q$, and strange quarks $s$ (listed
separately for 2+1 flavor models). We also list the maximum (TOV) mass
$M_{\rm TOV}$, the radius $R_{1.4}$ and dimensionless tidal deformability $\Lambda_{1.4}$
of a $1.4\,M_\odot$ neutron star, and any Refs. For BSK16 and RATP $R_{1.4}$ and $\Lambda_{1.4}$ are not given as, as for a 1.4 $M_{\odot}$ NS causality is broken $c_s>1$ which we do not consider.}
\label{tab:eos_summary}
\end{longtable}
\endgroup
\twocolumngrid
\bibliography{references.bib}

\end{document}